\documentclass[acmsmall, screen]{acmart}

\usepackage{graphicx}
\usepackage{textcomp}
\usepackage[table,xcdraw]{xcolor}
\usepackage{xkcdcolors}
\usepackage{tikz}
\usetikzlibrary{shapes.misc}
\usepackage{wrapfig}
\usepackage{subcaption}

\usetikzlibrary{decorations.pathreplacing,calc}
\newcommand{\tikzmark}[1]{\tikz[overlay,remember picture] \node (#1) {};}

\newcommand*{\AddNote}[4]{%
    \begin{tikzpicture}[overlay, remember picture]
   \draw [decoration={brace,amplitude=0.5em},decorate,ultra thick,black]
  ($(#3)!(#1.north)!($(#3)-(0,1)$)$) --  
  ($(#3)!(#2.south)!($(#3)-(0,1)$)$)
 node [align=center, text width=1cm, pos=0.5, anchor=west] {#4};
    \end{tikzpicture}
}

\usepackage{wrapfig}
\usepackage{listings}
\usepackage{xspace}
\usepackage{booktabs}
\usepackage{multirow}
\usepackage{hyperref}
\usepackage[linesnumbered,ruled,vlined]{algorithm2e}
\usepackage{adjustbox}
\usepackage{tcolorbox}
\definecolor{javared}{rgb}{0.6,0,0} 
\definecolor{javagreen}{rgb}{0.25,0.5,0.35} 
\definecolor{javapurple}{rgb}{0.5,0,0.35} 
\definecolor{javadocblue}{rgb}{0.25,0.35,0.75} 
\definecolor{diffincl}{HTML}{00A64F}
\definecolor{diffrem}{HTML}{ED135A}
\definecolor{highlightred}{HTML}{e99040}

\tcbset{
  mystyle/.style={
    colback=xkcdLightGrey,
    colframe=black,
    boxrule=0.9pt,
    boxsep=1pt,
    left=3pt,right=4pt,
    top=3pt,bottom=4pt,
    before skip=6pt,
    after skip=8pt,
    width=\linewidth
  }
}

\newcommand{\tool}{\textsc{Cascade}\xspace}

\newcommand{\circled}[1]{\tikz[baseline=(char.base)]{\node[shape=circle,draw,inner sep=1pt] (char) {#1};}}

\setcopyright{cc}
\setcctype{by}
\acmDOI{10.1145/3808175}
\acmYear{2026}
\acmJournal{PACMSE}
\acmVolume{3}
\acmNumber{FSE}
\acmArticle{FSE168}
\acmMonth{7}
\acmSubmissionID{fse26mainb-p1294-p}
\received{2026-02-25}
\received[accepted]{2026-03-24}

\begin{document}
\title{\tool: Detecting Inconsistencies between Code and Documentation with Automatic Test Generation}

\author{Tobias Kiecker}
\orcid{0009-0003-4744-0900}
\affiliation{%
  \institution{Humboldt-Universität zu Berlin}
  \city{Berlin}
  \country{Germany}
}
\email{tobias.kiecker@hu-berlin.de}

\author{Jan Arne Sparka}
\orcid{0000-0002-5886-4595}
\affiliation{%
  \institution{Humboldt-Universität zu Berlin}
  \city{Berlin}
  \country{Germany}
}
\email{sparkaja@hu-berlin.de}

\author{Martin Reuter}
\orcid{0009-0001-5435-3077}
\affiliation{%
  \institution{Humboldt-Universität zu Berlin}
  \city{Berlin}
  \country{Germany}
}
\email{reuterma@student.hu-berlin.de}

\author{Albert Ziegler}
\orcid{0000-0003-0937-7464}
\affiliation{%
  \institution{XBOW}
  \city{Uppsala}
  \country{Sweden}
}
\email{albert@xbow.com}

\author{Lars Grunske}
\orcid{0000-0002-8747-3745}
\affiliation{%
  \institution{Humboldt-Universität zu Berlin}
  \city{Berlin}
  \country{Germany}
}
\email{grunske@informatik.hu-berlin.de}

\begin{abstract}
Maintaining consistency between code and documentation is a crucial yet frequently overlooked aspect of software development. Even minor mismatches can confuse API users, introduce new bugs, and increase overall maintenance effort. 
This creates demand for automated solutions that can assist developers in identifying code-documentation inconsistencies. However, since automatic reports still require human confirmation, false positives carry serious consequences: wasting developer time and discouraging practical adoption.

We introduce \tool (\textbf{C}onsistency \textbf{A}nalysis for \textbf{S}ource \textbf{C}ode \textbf{A}nd \textbf{D}ocumentation through \textbf{E}xecution), a novel tool for detecting inconsistencies with a strong emphasis on reducing false positives. 
\tool leverages Large Language Models (LLMs) to generate unit tests directly from natural-language documentation. 
Since these tests are derived from the documentation, any failure during execution indicates a potential mismatch between the documented and actual behavior of the code.
To minimize false positives, \tool also generates code from the documentation to cross-check the generated tests. By design, an inconsistency is reported only when two conditions are met: the existing code fails a test, while the code generated from the documentation passes the same test.

We evaluated \tool on a novel dataset of 71 inconsistent and 814 consistent code-documentation pairs drawn from open-source Java projects. Further, we applied \tool to additional Java, C\#, and Rust repositories, where we uncovered 13 previously unknown inconsistencies, of which 10 have subsequently been fixed, demonstrating both \tool's precision and its applicability to real-world codebases.
\end{abstract}



\maketitle

\section{Introduction}
Inconsistencies between code and its documentation are a long-standing issue in software development~\cite{rani2023decade, tan2007icomment, reiss2006incremental}. A common oversight for developers is to update code without corresponding updates of the documentation~\cite{ratol2017detecting}. 
This can hinder the efficient use of libraries~\cite{uddin2015api}, increase the time required to understand a program~\cite{xia2017measuring}, or decrease the maintainability of software~\cite{lenhard2019exploring, de2005study}. Inconsistent 
documentation can also lead to the introduction of bugs in other parts of the program later in the development process~\cite{zhong2009inferring}. Although good documentation demonstrably benefits developers~\cite{xia2017measuring}, the act of keeping it in sync with evolving source code is repeatedly neglected~\cite{summaryzhu2019automatic}. 

\paragraph*{Example} Figure~\ref{fig:example} shows an example of an inconsistency in terms of a semantic mismatch between documentation and code. The \texttt{startsWithAny} string utility method from Apache's \texttt{commons-lang} checks whether a given string starts with any string from a supplied list. The documentation erroneously claims that the method is case-insensitive. So, for example, it suggests that \texttt{startsWithAny("HELLO World!", \{"hell"\})} would return \texttt{true}. In reality, the method's implementation is case-sensitive, a fact later confirmed by the developers in a commit\footnote{\href{https://github.com/apache/commons-lang/commit/d1a3255600da34f4b69dc082c4441ae140452fee}{github.com/apache/commons-lang/commit/d1a325560...}} that changed only this single word in the Javadoc. The inconsistency likely arose from a copy-paste error involving other case-insensitive methods in the same class. The mistake is difficult to detect through manual inspection, as the actual string comparison occurs several layers deeper in the call hierarchy in the \texttt{startsWith} method (invoked in \textit{line 12}).
Static approaches that just have the code and documentation as ground truth and do not execute the method, e.g., with a test case, would miss the problem. To detect these kinds of inconsistencies dynamic analysis is needed.

\begin{figure}
\centering
\begin{lstlisting}[language=Java, escapechar=§]
/**
* Check if a CharSequence starts with any of an array of specified strings.
§[...Shortened for better display...]§
* @return §\textcolor{javadocblue}{\textit{true}}§ if the CharSequence starts with any of the the prefixes, §\colorbox{highlightred}{case insensitive}§,
* or both §\textcolor{javadocblue}{\textit{null}}§ §[...]§
*/
public static boolean startsWithAny(final CharSequence string, final CharSequence... searchStrings){
   if (isEmpty(string) || ArrayUtils.isEmpty(searchStrings)){
      return false;
   }
   for (final CharSequence searchString : searchStrings){
      if (startsWith(string, searchString)) {
         return true;
   }  }
   return false;
}

    \end{lstlisting}
    
    \caption{
    Method \texttt{startsWithAny} from the \texttt{StringUtils} class in apache commons-lang. The docstring claims that the string matching is case-insensitive, while in reality it is case-sensitive.
    }
    \label{fig:example}
    \Description{a code example}
\end{figure}

\begin{figure*}[htp]
    \centering
    \includegraphics[width=0.8\linewidth]{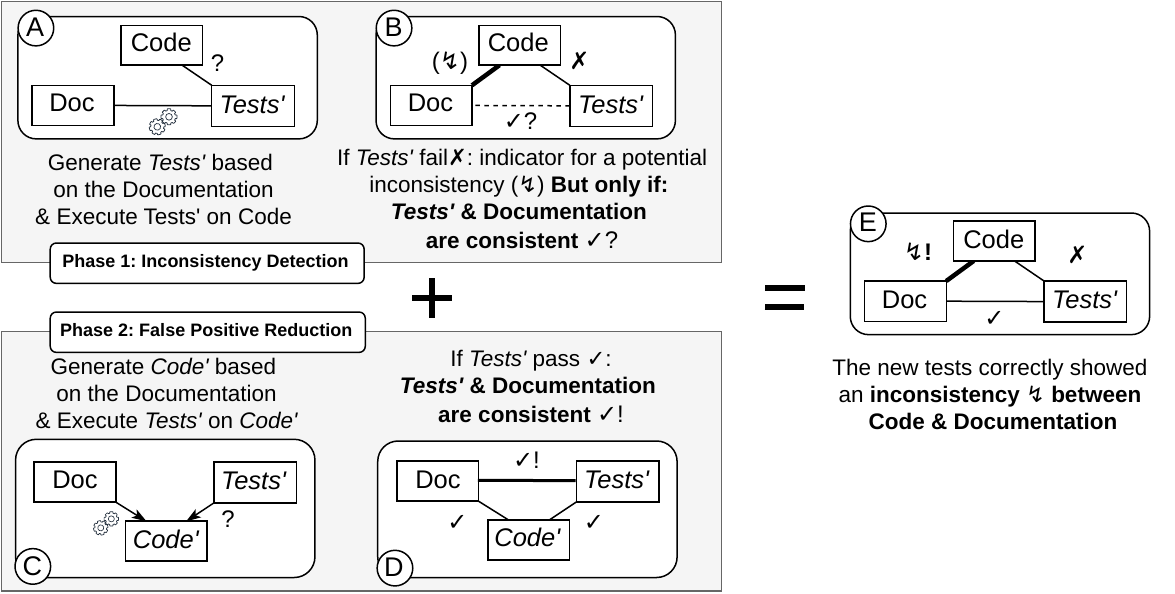}
    \caption{Intuition behind \tool. Phase 1 detects potential inconsistencies under the assumption that tests and documentation are aligned, and Phase 2 confirms this alignment.}
    \Description{This is the basic idea of our approach it consists of 4 parts. Step 1: Are documentation and code consistent? Step 2: Generate new tests consistent with the documentation. Step 3: Execute new tests against the code. Step 4: if the new tests failed we found an indicator for an inconsistency between documentation and code}
    \label{fig:idea}
\end{figure*}

\medskip
Early work in detecting inconsistencies relied on specific language compiler features~\cite{inconR}, rule-based heuristics~\cite{ratol2017detecting} or older Natural Language Processing (NLP) techniques~\cite{tan2012tcomment}.
With the increasing adoption of machine learning across all NLP tasks~\cite{MLinNLP}, recent work has also applied it, for example in the deep-learning classifier by Panthaplackel et al.~\cite{panthaplackel2021deep}. 

LLMs have recently shown an unprecedented ability to work with natural language, and this ability has been harnessed for many different software engineering applications~\cite{tian2024debugbench, lo2023trustworthy, jimenez2023swe}. While also promising for the inconsistency detection task, 
these models are notoriously unreliable. Minor changes in prompt wording can lead to divergent~\cite{ye2023assessing} or outright hallucinated outputs~\cite{huang2025survey}.
They sometimes even lack internal consistency~\cite{ye2023assessing}: asking an LLM a semantically similar question in a different way can produce largely different results.

\paragraph*{Example} During our experiments, we asked an LLM (\texttt{gpt-4.1-mini}\footnote{\href{https://platform.openai.com/docs/models/gpt-4.1-mini}{platform.openai.com/docs/models/gpt-4.1-mini}}) whether the example in Figure~\ref{fig:example} contained an inconsistency. Most of the time, it returned a clear \textit{no}, completely missing the point in its explanation and focusing only on the general structure and the usage examples in the documentation. When asked the semantically similar, but oppositely framed, question of whether the method and documentation were consistent, it sometimes returned a \textit{no} and in one cases gave the correct explanation by (correctly) guessing that the underlying \texttt{startsWith} method was case-sensitive. 
This highlights the volatility that pure LLM approaches often show. 
To conclude, traditional techniques~\cite{tan2012tcomment, jdoctor, tan2007icomment} are narrowly scoped and LLM-based methods~\cite{carlama, docchecker, sungmin} are highly volatile. Both shortcomings increase the risk of false alarms, a common problem in static analysis.

Static analysis tools, which are designed to detect potential bugs in software, frequently generate high rates of false positives~\cite{shen2011efindbugs}. This inefficiency can lead to a reluctance to adopt such tools~\cite{croft2021empirical} due to the time wasted on investigating false positives~\cite{kharkar2022learning}. Developers are more likely to engage with tools that are perceived to provide actionable insights with minimal noise~\cite{tomasdottir2018adoption}. Krishnamurthy et al.~\cite{krishnamurthy2019scientific} supports the claim that high false positive rates can deter even experienced developers from utilizing static analysis tools. Given this evidence, we adopt a precision‑first stance: minimizing false positives is the primary goal, even at the cost of recall.

\paragraph{Tool} We introduce \tool (\textbf{C}onsistency \textbf{A}nalysis for \textbf{S}ource \textbf{C}ode \textbf{A}nd \textbf{D}ocumentation through \textbf{E}xecution) to address the challenge of identifying general code-documentation inconsistencies without triggering excessive false positives. 
We prioritize precision over recall. \tool is intended to flag only inconsistencies that are strongly supported by evidence, accepting that some true inconsistencies might be missed.
The high-level idea of \tool is shown in Figure~\ref{fig:idea}.
\circled{A}
\tool leverages the natural language understanding capabilities of LLMs to automatically generate executable unit tests including oracles from method-level documentation (e.g., Javadoc). These tests aim to capture the described behavior.  
\circled{B} 
If at least one generated test case fails when executed on the code, this strongly suggests a mismatch between documentation and implementation. If the goal is high recall, stopping at this point is acceptable. However, LLM's are imperfect and can hallucinate or misinterpret intent and thus generate wrong test cases. To mitigate this risk, \tool performs a second validation phase.
\circled{C}
The same documentation is used to synthesize a new implementation of the method. 
This synthesized version is expected to pass the documentation-derived tests if both were correctly generated and aligned with the documented behavior.
\circled{D}
We assume that LLM-based code generation is typically more reliable than test generation~\cite{ni2023lever, chen2022codet}, and that the joint success of independently generated code and tests on the same specification increases confidence that both are correct representations of the documentation.
\circled{E}
Thus, an inconsistency is flagged only when the following two conditions are met: (1) the original implementation fails a generated test, and (2) the synthesized implementation passes the same test. This dual check design significantly reduces false positives that could arise from LLM errors or ambiguities in the documentation.

\paragraph*{Dataset}
To enable a meaningful evaluation of \tool, we constructed a novel dataset of method-documentation pairs that is, to our knowledge, the first to consist of real-world confirmed semantic inconsistencies within executable Java projects. Specifically, we mined and examined commits where developers explicitly corrected inaccurate documentation, and we verified that both the original and corrected code compile successfully. This approach ensures that the dataset reflects authentic, developer-acknowledged semantic code-documentation mismatches in real software projects. To improve generalizability and better reflect realistic class distributions, we additionally collected approximately ten times as many consistent cases as inconsistent ones.
In contrast, existing code-documentation inconsistency datasets fall short in several ways. Non-human-curated datasets that are mined based on heuristics, such as those by Panthaplackel et al.~\cite{panthaplackel2020learning}, tend to label superficial issues (e.g., typos or minor wording changes) as ``inconsistent'' and lack executable code or context information. Lee et 
al.~\cite{metamon} used a mutation-based dataset, that relies on synthetic error injection, while others generate documentation with LLMs~\cite{sungmin}, potentially misrepresenting natural developer behavior. These limitations make our dataset uniquely suited for robust and realistic evaluation of tools like \tool.

Overall, the main contributions of this paper are:
\begin{itemize}
    \item a novel approach and tool \tool that detects using newly generated test cases to uncover inconsistencies between code and documentation and newly generated code to prune false positives (Section~\ref{sec:cascade});
    \item a manually curated dataset of commits from executable Java projects that have a known inconsistency between code and documentation (Section~\ref{sec:dataset});
    \item an evaluation of \tool on this new dataset and on real-world projects of different programming languages (Section~\ref{sec:eval}), where we found previously unknown and confirmed inconsistencies.
\end{itemize}

\section{Related Work}\label{sec:rel}
A key aspect of high-quality documentation is its consistency with the code it describes~\cite{rani2023decade}. There are two types of approaches to ensure this: preventing the introduction of inconsistencies into the codebase (just-in-time) and detecting already existing inconsistencies (post-hoc).

\subsection{Avoiding Inconsistencies}
Just-in-time (JIT) techniques aim to update documentation at the time of code changes. Panthaplackel et al.~\cite{panthaplackel2020learning, panthaplackel2021deep} proposed machine learning models to predict and classify comment updates based on code diffs, and Rong et al. fine-tuned a code LLM for the same purpose~\cite{carlama}. Liu et al.~\cite{liu2020automating, liu2021just} developed seq-to-seq models for generating updated comments from code history. Stulova et al.'s upDoc~\cite{upDoc} maintains code-documentation mapping before and after changes to highlight potential mismatches. \tool aims at detecting existing inconsistencies, thus these approaches ~\cite{panthaplackel2020learning, panthaplackel2021deep,liu2020automating, liu2021just,upDoc} are only adjacently related.

\subsection{Detecting Inconsistencies}
Ensuring good consistency in a project can also be achieved by analyzing projects in what is referred to as a post-hoc setting~\cite{panthaplackel2021deep}. It is generally not advisable to automatically fix inconsistencies that are found, as developers, particularly in the context of AI tools, often prefer tools that offer hints or advice rather than perform actions without oversight~\cite{developersWant}. Therefore, most research in this area focuses on detecting inconsistencies rather than resolving them. This is the context in which our approach, \tool, is situated. There have been several other notable approaches in this domain. 

Early research on code-documentation consistency centered on handcrafted rules and static analysis.  
Ouyang and Hua's tool $'R$~\cite{inconR} exploits Rust's built-in documentation tests, offering precise checks at the cost of language flexibility.  
Similarly, Sondhi and Purandare~\cite{sondhi2019segate} match API documentation against a curated database of regular-expression templates for string operations, while Ratol et al.~\cite{ratol2017detecting} flag parameter mismatches and unit-conversion errors through custom heuristics.  
Tan et al.'s \textsc{iComment}~\cite{tan2007icomment} generalizes the idea by mining logical constraints from C comments and statically verifying them.

To go beyond purely static checks, several systems generate or use concrete executables.  
As a direct update to \textsc{iComment}, Tan et al. proposed \textsc{@tComment}~\cite{tan2012tcomment}, which extracts rules regarding \texttt{null}-values from Java comments and synthesizes test cases with Randoop~\cite{randoop} to expose violations.  
\textsc{JDoctor}~\cite{jdoctor} adopts the same strategy for Javadoc pre- and post-conditions, it is also limited to \texttt{@param}, \texttt{@return} and \texttt{@throws} tags.  
More recently, Panthaplackel et al.~\cite{panthaplackel2020learning} train neural classifiers to detect mismatches between comments and methods~\cite{panthaplackel2021deep} and Dau et al. built \textsc{DocChecker}~\cite{docchecker} to automatically handle the extraction and classification on top of that model; Gao et al.~\cite{gao2021automating} learn to remove resolved `\textit{TODO}' annotations; and Kang et al.~\cite{sungmin} focus on the correctness of LLM generated docstrings~\cite{sungmin}.  
While these learning-based approaches generalize better than rule-based ones, they still rely either on very specific features or on large labelled corpora.

LLM-prompting has opened a new direction.  
Lee et al.'s \textsc{Metamon}~\cite{metamon} asks an LLM whether regression tests generated by EvoSuite~\cite{evosuite} are consistent with the documentation.
However, their evaluation is also confined to Java methods with \texttt{@param} and \texttt{@return} tags~\cite{metamon}.  
Zhang et al.'s RustC\textsuperscript{4} combines LLM reasoning with Rust-specific static checks~\cite{rustc4}. And Rong et al.~\cite{carlama} fine-tuned a CodeLLaMA model for Javadoc inconsistencies. 
Despite promising recall, LLM-based detectors suffer from volatility: minor prompt changes or paraphrases can yield contradictory answers~\cite{ye2023assessing, huang2025survey}, inflating false positive rates. 

In contrast to these approaches, \tool is designed to operate on all facets of code documentation -- compared to template-based detectors~\cite{sondhi2019segate}, Javadoc-tag specific approaches \cite{tan2012tcomment, jdoctor, metamon, carlama} or specialized `\textit{TODO}'-comment removal~\cite{gao2021automating}. It employs targeted, LLM-guided test-case generation -- instead of the untargeted or evolutionary regression test generation used by \textsc{@tComment}~\cite{tan2012tcomment}, \textsc{JDoctor}~\cite{jdoctor} or \textsc{Metamon}~\cite{metamon}). 
Besides these technical differences, \tool's primary goal is further to \emph{significantly reduce false positives}, making post-hoc inconsistency detection practical for day-to-day development.

\section{\tool}\label{sec:cascade}
For each extracted function, \tool predicts whether its documentation is consistent with the implemented code, and outputs one of two possible results:
\textbf{positive} - an inconsistency is very likely or \textbf{negative} - we did not find sufficient evidence for an inconsistency). This entire procedure consists of two main phases, as described in Algorithm~\ref{alg:cascade}.

\paragraph{\textbf{Phase 1: Inconsistency Detection}}

We first generate new Tests $\mathcal{T}$ (\textit{line 2}). 
In principle, the generation part of the analysis could use any test synthesis approach that is able to run using only documentation as the source for the generation. Specifically in this implementation, we use Large Language Models (LLMs) as generators, as these are currently the best approaches to generate code and tests based on natural language descriptions~\cite{schafer2023empirical, ni2023lever, wang2021codet5}.


\begin{wrapfigure}[40]{r}{0.51\textwidth}
\centering
\begin{algorithm}[H]
\footnotesize
\caption{DetectInconsistency($\mathit{doc},\mathit{code}$)}\label{alg:cascade}
\LinesNumbered
\KwIn{$\mathit{doc}$ --- documentation of a function}  
\KwIn{$\mathit{code}$ --- the function's current implementation}
\KwOut{verdict --- \textsf{Positive} (inconsistency) or \textsf{Negative}\tikzmark{right}}
\SetKwFunction{DetectInconsistency}{DetectInconsistency}
\SetKwFunction{GenTests}{GenerateTests}
\SetKwFunction{Exec}{ExecuteTests}
\SetKwFunction{Repair}{Repair}
\SetKwFunction{GenCode}{GenerateNewCode}
\SetKwProg{Fn}{Function}{:}{}
\tikzmark{left}\Fn{\DetectInconsistency{$\mathit{doc}, \mathit{code}$}}{
    $\mathcal{T} \leftarrow$ \GenTests{$\mathit{doc}$}\tikzmark{p1top}

    \tcp*[h]{Execute new tests on original code}
    
    $\textit{res}_1 \leftarrow$ \Exec{$\mathcal{T},\mathit{code}$}

    \If{\textsf{CompilerError} $\in res_{1}$ }{
    
        \For{$i \leftarrow 1$ \KwTo $3$}{
        
            $\mathcal{T} \leftarrow$ \Repair{$\mathcal{T}, \textsf{CompilerError}$}
            
            $\textit{res}_1 \leftarrow$ \Exec{$\mathcal{T},\mathit{code}$}

             \If{\textsf{CompilerError} $\notin res_{1}$}{\textbf{break}}
        }

        \If{\textsf{CompilerError} $\in res_{1}$ }{
        
            \tcp*[h]{uncompilable after 3 tries}
            
            \Return \textsf{Negative}
        }
    }

    \If{$\textit{res}_1.failed = \varnothing$ }{
        \Return \textsf{Negative}\tikzmark{p1bottom}
    }
    
    $\mathit{code}' \leftarrow$ \GenCode{$\mathit{doc}$}\tikzmark{p2top}
    
    \tcp*[h]{re-run tests on new code}
    
    $\textit{res}_2 \leftarrow$ \Exec{$\mathcal{T},\mathit{code}'$}
    
    \tcp*[h]{classify test-outcome transition}
    
    $\mathsf{p2p}\gets0,\;\mathsf{p2f}\gets0,\;\mathsf{f2p}\gets0,\;\mathsf{f2f}\gets0$

    \ForEach{$t\in\mathcal{T}$}{
        \uIf{$t \in \textit{res}_1\textit{.passed} \wedge t \in \textit{res}_2\textit{.passed} $}{
            $\mathsf{p2p} \mathrel{{+}{=}} 1$
        }
        \uElseIf{$t \in \textit{res}_1\textit{.passed} \wedge t \in \textit{res}_2\textit{.failed} $}{
            $\mathsf{p2f} \mathrel{{+}{=}} 1$
        }
        \uElseIf{$t \in \textit{res}_1\textit{.failed} \wedge t \in \textit{res}_2\textit{.passed} $}{
            $\mathsf{f2p} \mathrel{{+}{=}} 1$
        }
        \Else{
            $\mathsf{f2f} \mathrel{{+}{=}} 1$
        }
    }

    \uIf{$\mathsf{f2p} > 0$ \textbf{and} $\mathsf{p2f} = 0$}{

        \tcp*[h]{Detected inconsistency}
        
        \Return \textsf{Positive} 
    }
    \Else{

        \tcp*[h]{No inconsistency found}
    
        \Return \textsf{Negative}\tikzmark{p2bottom}
    }}
\AddNote{p1top}{p1bottom}{right}{\rotatebox[origin=c]{270}{\textbf{Phase 1: Inconsistency Detection}}}
\AddNote{p2top}{p2bottom}{right}{\rotatebox[origin=c]{270}{\textbf{Phase 2: False Postive Reduction}}}
\end{algorithm}
\end{wrapfigure}

Several existing text-to-testcase approaches were not suitable under our constraints: they either do not target Java~\cite{pytester}, do not directly generate executable unit tests~\cite{jdoctor}, or take natural language artefacts other than function-level documentation as input~\cite{boddu2004, carvalho2014, fischbach2023}. 
Although Alagarsamy et al.~\cite{text2test} address text-to-testcase generation, their approach is tightly coupled to a task-specific evaluation setup rather than a reusable, modular component. 
Another closely related approach by Kang et al.~\cite{sungmin} does not provide any available implementation; however, our test generation component is loosely inspired by their prompting strategy.

For any given function under tests, we first prompt the LLM to generate a list of testable behavior from the documentation. These behaviors are usually returned in the form of `\textit{if [condition] then [behavior]}'. E.g., if the input parameters have certain values, the method throws a specific exception or if some other method has been called before it should return a positive value. 
From this list, we create a test-class template with one test stub per item. Each method includes a comment describing the intended behavior.
Finally, we prompt the LLM to complete the tests, providing the relevant class context (constructors, fields, other methods, etc.). The full multistage prompting is available in our supplementary material.\footnote{\href{https://github.com/TobiasKiecker/CASCADE/blob/FSE2026-v1.0.0/src/cascade/generation/test/MultiStepJavaTestGenerator.py}{github.com/TobiasKiecker/CASCADE/.../test/MultiStepJavaTestGenerator.py}}

Next, we execute this new test class on the original project and collect a list of passed and failed tests (\textit{line 3}).
If the new test class does not compile, we automatically parse the compiler errors and ask the LLM again in a type of repair loop to fix them (\textit{lines 4-9}). If the repair step is attempted three times and the tests still do not compile, we predict a \textbf{negative} (No inconsistency) (\textit{line 11}) since we have not found any evidence for an inconsistency.

We are left with two possible outcomes: 

\emph{1. All generated tests pass} 
If no generated test fails, we have no indicator for an inconsistency. Note that, this outcome does not strictly confirm consistency, as the discriminating test case might simply be missing, due to imperfect generation. While there's also a slight chance that some tests pass trivially (e.g., tautologies), a manual review suggests that this is very rare. Thus, we predict a \textbf{negative} (no inconsistency) and terminate the run (\textit{line 13}).

\emph{2. At least one generated test fails} 
A failing test can indicate an inconsistency. However, test generation from documentation can be flawed, and thus it is possible that this generated test case is just wrong. The LLM essentially has to solve the oracle problem for these methods and could generate tests that are either always wrong or do not adhere to the documentation at all. 
Therefore, we can only predict a positive with reservations and continue with the verification step to reduce the amount of false positives.

\paragraph*{Example} 
For our running example in Figure~\ref{fig:example} this stage might produce a failing test case such as: \texttt{assertTrue(startsWithAny("HELLO world!", {"hello"}))} if the method was truly case-insensitive like described in the documentation this test should have passed. 
However, this only holds true if the LLM extracted the correct information from the documentation, thus further verification of the test case is needed. 

\paragraph{\textbf{Phase 2: False Positive Reduction}}
To confirm that the potential inconsistency is real, we enter Phase~2 and generate a new implementation of the documentation: $code'$ (\textit{line 14}). 

Prior work shows that general code generation is usually more reliable than direct test generation~\cite{ni2023lever, chen2022codet}, thus this step filters out broken tests.
If a new test fails on both the regenerated code as well as on the original, we cannot trust it.

The code-generation prompt is simpler than the test-generation prompt: we provide the original class with the body of the method removed, retaining its signature and documentation, and ask the LLM to implement the missing method. 
The exact prompt can be found with the supplementary material and in the code.

We then run all generated tests from Phase 1 again on this newly generated code (\textit{line 15}) and collect the individual test results. 
Combining the results of both phases, we can classify each individual test case based on its two outcomes (\textit{lines 16-25}):

\begin{itemize}
    \item[\textbf{p2p}] pass to pass - These test cases passed both on the original and on the new code. These can be safely ignored. Either they are trivial or they do test some part of the documentation that is consistent with both the old code and the new. 
    \item[\textbf{f2f}] fail to fail - Tests that failed on both the original code as well as on the new code. These were most likely generated wrong. We acknowledge that there are other cases e.g. there is an edge case that is wrong in the old code and also not properly implemented in the new code, but in this case, the test case is not helpful to us either. So, we also ignore them.
    \item[\textbf{f2p}] fail to pass - Test cases that fail on the original code but pass on the documentation aligned new code, are the actual indicators for inconsistencies that we are looking for. These tests show behavior that is described in the documentation (as shown by two independently generated artifacts), but not reflected in the code.
    \item[\textbf{p2f}] pass to fail - These are interesting cases that passed on the original code but now failed on the new code. These can actually give us an indicator for wrongly generated code. It shows that the new code is missing at least some documentation-consistent functionality that the original code had. Thus, if such a case exists, we cannot trust the new code to actually follow the new documentation and accordingly, also not trust the f2p test cases to really show an inconsistency. We disregard the entire function and predict a negative instead. 
\end{itemize}

In summary, we finally predict a \textbf{positive} result (an inconsistency) if there are no test cases that changed from pass on the original code to fail on the regenerated code (p2f) and at least one test case that changes from fail to pass (f2p) (\textit{line 27}). Under these conditions, we have simultaneous evidence that, on one hand, the regenerated code is not generated incorrectly (all retained tests pass) and on the other hand, the f2p test exposed a mismatch in the original implementation. Hence, we can confidently conclude that a genuine inconsistency exists.
\paragraph*{Example} Assuming, for our running example in Figure~\ref{fig:example}, a wrong test case would have been generated: \texttt{assertFalse(startsWithAny("Hello world!" , \{"h" , "he", "world"\}))} because the test generation LLM made the mistake of thinking the target string has to start with all of the strings in the list. This test would fail on the original code. But it would also fail on the new generated code and thus we knew not to trust this test case (f2f).
On the other hand, the correct (failing) test case from before \texttt{assertTrue(startsWithAny("HELLO world!", \{"hell"\})} would pass on the new code and we can predict an inconsistency with high certainty (f2p). 
If for example, a test case like: \texttt{assertFalse(startsWithAny("Hello" , \{"a"\})} would suddenly fail on the new code we would know that something went wrong during code generation (e.g., the new code might just always return \texttt{True}) and we cannot trust the test case as a predictor (p2f). 

\tool is a post-hoc approach and can be used on a project to find existing inconsistencies. It could be used alongside other test suites overnight, leaving a human developer to evaluate the results. This is why the goal to reduce the False Positives is so important to decrease the burden on a developer.

\section{Dataset}\label{sec:dataset}
\subsection{Previous Datasets}
Panthaplackel et al.~\cite{panthaplackel2020learning} provide one of the largest dataset in the area, which has been used for several approaches.~\cite{panthaplackel2021deep, docchecker, carlama}, mostly to train classical machine learning classifiers. It contains pairs of functions and Javadoc comments extracted from historical commits, but it offers neither project context nor the exact commit hashes from which the methods were taken. As a result, it is not possible to compile and run any new tests on these projects.  In addition, the ``inconsistency'' label is inferred from a very simple heuristic: commits that modify only documentation, which frequently captures trivial typo fixes or wording changes. Without ground-truth confirmation by the developers, the labels
remain unreliable.

Lee et al.~\cite{metamon} used Defects4J~\cite{defects4j} as a base for their evaluation. Defects4J was designed for bug detection research, not documentation analysis. Assuming every buggy version violates the documentation is often untrue: many bugs change internal edge cases while leaving the API contract intact. They compound the issue by injecting artificial mutants, producing code that is supposedly no longer following its documentation. This can lead to nonsensical code that would never occur in the real world. In addition, the well-studied Defects4J bugs are widely discussed and described in literature,
so LLMs may have already memorized them, making any comparison with LLM-based approaches inherently flawed.

Kang et al.~\cite{sungmin} focused on the consistency of documentation generated by LLMs rather than written by developers. While useful for studying hallucination and the ability of LLMs to generate good documentation, it diverges from the diverse, sometimes messy style of real-world comments and therefore would be an unrealistic benchmark for \tool.

Wen et al.~\cite{wen2019large} tracked Java commits that change comments and manually labeled a subset of them. Many of their samples include parts where a new comment was added or outright false positives~\cite{upDoc}.
However, we did use their taxonomy of code-documentation changes to guide the inclusion and exclusion criteria for our dataset.

\subsection{Our Dataset}
\begin{table}[tb]
\centering
  \caption{Dataset Collection Statistics.}\label{tab:dataset}
\begin{tabular}{@{}lrrcrrc@{\hskip 4pt}rr@{}}
\toprule
\multirow{2}{*}{author/project}  & \multicolumn{2}{c}{\#commits} &           & \multicolumn{2}{c}{\#inconsistent func.} &           & \multicolumn{2}{c}{{\#consistent func.}} \\ \cmidrule(lr){2-3} \cmidrule(lr){5-6} \cmidrule(lr){8-9}
                                 & match      & exam.      &           & found              & runable             &           &  paired  &  extra              \\ \midrule
apache/commons-lang              & $\sim$1.1k    & 250           &           & 11                 & 10                  &           & \textit{10} & \textit{176}             \\
{apache/commons-compress}          & {973}           & {250}           &           & {8}                  & {3}                   &           & \textit{3} & \textit{9}                \\
apache/commons-math              & 789           & 250           &           & 16                 & 13                  &           & \textit{13} & \textit{163}             \\
apache/commons-collections       & 631           & 250           &           & 5                  & 2                   &           & \textit{2} & \textit{25}               \\
{apache/commons-codec}             & {454}           & {250}           &           & {6}                  & {4}                   &           & \textit{4} & \textit{24}               \\
assertj/assertj                  & 446           & 250           &           & 4                  & 2                   &           & \textit{2} & \textit{1}                \\
google/guava                     & 394           & 250           &           & 2                  & 2                   &           & \textit{2} & \textit{24}               \\
{jfree/jfreechart}                 & {281}           & {250}           &           & {3}                  & {1}                   &           & \textit{1} & \textit{19}               \\
{apache/commons-csv}               & {244}           & {244}           &           & {8}                  & {4}                   &           & \textit{4} & \textit{10}               \\
{apache/commons-cli}               & {215}           & {215}           &           & {4}                  & {4}                   &           & \textit{4} & \textit{40}               \\
FasterXML/jackson-databind       & 212           & 212           &           & 3                  & 2                   &           & \textit{2} & \textit{16}               \\
{JodaOrg/joda-time}                & {160}           & {160}           &           & {10}                 & {8}                   &           & \textit{8} & \textit{80}               \\
{google/gson}                      & {145}           & {145}           &           & {4}                  & {3}                   &           & \textit{3} & \textit{27}               \\
{FasterXML/jackson-core}           & {145}           & {145}           &           & {4}                  & {2}                   &           & \textit{2} & \textit{23}               \\
{jhy/jsoup}                        & {97}            & {97}            &           & {8}                  & {8}                   &           & \textit{8} & \textit{80}               \\
graphhopper/jsprit               & 42            & 42            &           & 3                  & 3                   &           & \textit{3} & \textit{106}              \\
{\textit{Other considered Projects\footnotemark}} & {\textit{1978}} & {\textit{1214}} &  & {\textit{8}}         & {\textit{0}}          &  & {-}                           \\ \midrule
\textbf{Total}                   & 7206          & 4474          &           & 107                &\textbf{71}        &           & \textit{71} & \textit{743}             \\ \bottomrule
\end{tabular}
\end{table}
\footnotetext{Projects not included in the final dataset (examined numbers): graphstream/gs-core (27), jgrapht/jgraph (178), apache/logging-log4j2 (250), apache/storm (75), Multibit-Legacy/multibit-hd (13), mockito/mockito (250), awslabs/amazon-dynamodb-lock-client (1), apache/commons-jxpath (78), google/closure-compiler (53), FasterXML/jackson-dataformat-xml (28), junit-team/junit4	(195), qos-ch/logback (66)}
None of the reviewed datasets simultaneously provides (1) runnable source code, (2) documentation-level inconsistencies, and (3) developer-validated semantic mismatches. 
We therefore curated a new dataset of documentation-fix commits from large active Java projects. Every instance compiles, every inconsistency was acknowledged and corrected by the original author, and each fix warranted its own commit, yielding a clean, executable ground truth dataset.

\paragraph{Inconsistent Samples}
Statistics of our dataset collection can be seen in Table~\ref{tab:dataset}.
We started with the five largest projects used for evaluation by Blasi et al.~\cite{jdoctor}. 
Their goal was to generate specifications from documentation and they chose these projects for their focus on good documentation.
These projects were joined with ten other high-starred maven buildable GitHub projects and the projects featured in Defects4J~\cite{defects4j}
Now, for each of the 28 selected projects, we used GitHub Search to find commits containing the term `\textit{Javadoc}' (\textit{\#commits match in the table})
and manually examined every matching commit, up to a maximum of 250 per project (\textit{\#commits exam.}). In total, this amounted to 4,474 commits manually reviewed.
For each commit, we individually checked all changed methods to determine if only the Javadoc was changed and whether the change constitutes a fix for an inconsistency. If it did, we marked down that specific function in the previous commit as inconsistent. In this previous commit the fix was not applied yet, and we could be highly certain that at this point code and documentation had been considered inconsistent. 
We explicitly excluded the following types of changes:
\begin{itemize}
    \item Simple formatting changes and typos except for functionality-altering ones (e.g., ``case insensitive'' instead of ``case sensitive'' from Figure~\ref{fig:example}).
    \item Changes to links in the documentation (e.g., \texttt{@see} or \texttt{@link} tags).
    \item The addition of documentation where there was previously none.
    \item Simple renaming of parameters to fit the signature without any other substantial changes to their description.
\end{itemize}
We filtered out private methods and abstract classes, which are rarely tested directly, as well as constructors which often have their documentation integrated into the class-level documentation.
This left us with 107 methods across 55 commits from 16 projects \textit{\#inconsistent func. found} in Table~\ref{tab:dataset}).

With a commit date range spanning from 2004 to 2024, we covered 20 years of Java development, encompassing various Java versions and conventions, and different coding styles. Considerable effort went into making these old versions of projects executable so that we could run the newly generated tests. This was impossible for some of the commits if the whole project was in a broken state. 
All changes were made with great care to avoid introducing new errors or inconsistencies in the functions under test. Unfortunately, we were not able to get every project to compile and therefore had to exclude these functions. We thus were left with 71 manually checked inconsistencies from runnable projects that were included into the dataset (\textit{\#inconsistent func. runable} in Table~\ref{tab:dataset}).

\paragraph{Consistent Samples} Because we want to evaluate the effectiveness of inconsistency-detection tools, we also need consistent code and documentation to serve as ground truth negative examples. Our dataset-collection process automatically provides the developer-corrected version of each inconsistency (\#consistent func. paired in the table), since the commit we located explicitly contains the fix. This gives us a balanced dataset; for every function we have one version with correct documentation and one with incorrect documentation, as identified by the project's developers.

However, since real-world data is not nessecarly balanced, we additionally wanted to evaluate on a more realistic dataset. Therefore, we included further consistent cases by extracting all other methods from the same class that contained documentation and were not modified in the commit (with a maximum of 20 per class) resulting in an additionall 743 consistent functions (\#consitent func. extra in the table). The rationale is that if a dedicated fix was made to a specific documentation issue in that class, the remaining unchanged methods are likely consistent. This is a weaker heuristic than using explicitly fixed cases, but it provides a sufficiently realistic approximation.

Finally, we have a balanced core dataset consisting of \textbf{142 method-Javadoc pairs}, 71 with an inconsistency and 71 without. Furthermore, we include 743 additional methods without an inconsistency to enable evaluation under a more realistic data distribution.
We provide the dataset with the supplementary material.\footnote{\href{https://github.com/TobiasKiecker/CASCADE/blob/FSE2026-v1.0.0/PaperEvaluation/dataset.zip}{github.com/TobiasKiecker/CASCADE/.../dataset.zip}}

\section{Experimental Design and Setup}\label{sec:eval}
Our evaluation consists of three parts. First, we assess the performance of \tool on the novel dataset by comparing it against several LLM baselines and two state-of-the-art inconsistency classification tools. 
Second, we conduct an ablation study on the same dataset to analyze the contribution of each component and design choice within \tool. 
Finally, we investigate the applicability of \tool to real-world software projects written in various programming languages, and present examples of actual inconsistencies it uncovered.

\subsection{Research Questions}
This evaluation is guided by the following research questions:
\begin{itemize}
    \item \textbf{RQ1: Performance}: How does \tool compare to other inconsistency detection approaches in terms of classification metrics such as precision, recall, and specificity?
    \item \textbf{RQ2: Ablation}: How do the individual components and design decisions in \tool impact its performance in detecting code-documentation inconsistencies?
    \item \textbf{RQ3: Real-World Utility}: How effective is \tool at detecting inconsistencies in large-scale, real-world software projects?
\end{itemize}
These questions reflect the goals of our evaluation: to establish \tool's relative effectiveness (RQ1), to understand the contribution of its internal architecture (RQ2), and to validate its practical value beyond controlled datasets (RQ3).

\subsection{Experiments}
For the main experiment, we compare our approach against three baselines: an LLM-Baseline, \textsc{DocChecker}~\cite{docchecker} and \textit{C4RLLaMA}~\cite{carlama}. All LLMs used for RQ1 and RQ2, (in the baselines and within \tool) were run with a temperature of 0 using the same model: \texttt{gpt-4.1-mini}.\footnote{\href{https://platform.openai.com/docs/models/gpt-4.1-mini}{platform.openai.com/docs/models/gpt-4.1-mini}} 
For RQ3, we had several different versions of \tool running during its development mainly using\texttt{gpt-4o-mini}.\footnote{\href{https://platform.openai.com/docs/models/gpt-4o-mini}{platform.openai.com/docs/models/gpt-4o-mini}}

Several tools have been proposed to reveal inconsistencies between source code and its documentation, yet most cannot serve as practical baselines for our evaluation.
\\\textit{Language support} disqualifies RustC\textsuperscript{4}~\cite{rustc4} and $'R$~\cite{inconR}, which both target Rust specifically, and \textsc{Metamon}~\cite{metamon}, whose analysis pipeline is bound to Java~8 and JUnit~4 due to its reliance on EvoSuite~\cite{evosuite}.
\\\textit{Reproducibility} is another limiting factor: the approach of Kang et al.~\cite{sungmin} and RustC\textsuperscript{4}~\cite{rustc4} lack any published implementation. The supplementary material of \textsc{Metamon}~\cite{metamon} is tightly coupled to their own dataset evaluation, making adaptation infeasible.
\\Finally, \textit{scope} is an important aspect. \textsc{@tComment}~\cite{tan2012tcomment} focuses only on defects regarding \texttt{null} values, while \textsc{JDoctor}~\cite{jdoctor} and \textsc{Metamon}~\cite{metamon} focus only on \texttt{@param}, \texttt{@return}, and \texttt{@throws} tags and therefore would miss inconsistencies expressed elsewhere in the Javadoc block.
After applying these criteria, two candidates that (1) support modern Java, (2) are publicly available, and (3) claim general-purpose coverage remain: \textsc{DocChecker} and \textit{C4RLLaMA}.

\paragraph{\textsc{DocChecker}}
Dau et al.~\cite{docchecker} extend Panthaplackel et al.'s approach~\cite{panthaplackel2021deep} with a model trained from ground up specifically for code–comment inconsistency detection (post-hoc). To make \textsc{DocChecker} runnable, we re-trained their model from scratch and modified a few lines in their provided library code. 
We verified the implementation on their clean test split and even achieved better results in F1 and accuracy than those reported in the original paper~\cite{docchecker}. We made this retrained model available with the supplementary material.
\footnote{\href{https://github.com/TobiasKiecker/CASCADE/blob/FSE2026-v1.0.0/PaperEvaluation/drivers/DocChecker/pretrained_model/README.md}{github.com/TobiasKiecker/CASCADE/.../DocChecker/pretrained\_model}}
However, we discovered another minor issue. For Java, the tool does not automatically extract the Javadoc; instead, it uses the first comment inside the function as the target documentation. To give \textsc{DocChecker} a fair evaluation, we implemented two wrappers:
\textbf{\textsc{DocChecker}-block:} compresses the entire Javadoc into a single-line comment at the beginning of the function and run \textsc{DocChecker} as is.
\textbf{\textsc{DocChecker}-single:} processes the Javadoc line by line, separately placing each line into the first comment and invoking the predictor for each. If at least one prediction indicates an inconsistency, we mark the sample as predicted inconsistent.
Due to the training dataset consisting of single line comments only, this will result in a fairer evaluation.

\paragraph{\textit{C4RLLaMA}}
Rong et al.~\cite{carlama} fine-tuned a pretrained LLM (CodeLLaMA-7B) instead of training a model from ground up. We re-trained and re-evaluated the model using their provided setup and scripts and include the resulting weights in the supplementary material.\footnote{\href{https://github.com/TobiasKiecker/CASCADE/blob/FSE2026-v1.0.0/PaperEvaluation/drivers/C4RLLaMA/weights/README.md}{github.com/TobiasKiecker/CASCADE/.../C4RLLaMA/weights}}
To interface with Javadoc, we mirror the wrappers from \textsc{DocChecker}:
\textbf{\textit{C4RLLaMA}-block:} pass the complete Javadoc string as a block (stripped of comment markers).
\textbf{\textit{C4RLLaMA}-single:} invoke the predictor individually for each Javadoc line and mark the method inconsistent if any line is flagged.
This aligns with the underlying dataset, which contains single-line comment edits~\cite{panthaplackel2020learning}.

\paragraph{LLM Baselines}
Instead of finetuning a pretrained LLM, it is also possible to use a general purpose LLM directly.
We define several LLM-based baselines for detecting inconsistencies between code and documentation, 
Notably, LLMs tend to exhibit inconsistency in their own outputs and are prone to hallucinations, e.g., when guessing what unfamiliar function calls within a method under test do. 
To systematically assess and mitigate this unreliability, we query the LLM in four distinct single-call settings. Two of these focus on direct questions with limited context: (1) asking whether the method and its documentation are consistent (LLM-S$_c$), and (2) asking whether there is an inconsistency between them (LLM-S$_i$). These are very close semantically but phrased inversely. One would expect that a method flagged consistent by the first call would not have an inconsistency found in the second, and vice versa. 
The other two baselines repeat the same binary questions but include more advanced contextual information (LLM-A$_c$ and LLM-A$_i$).  This extra context consists of the full class in which the method is defined, along with its associated documentation and other methods. This is the same information that we use in \tool for the test generation. This results in four single-query baselines (LLM-S$_c$, LLM-S$_i$, LLM-A$_c$ and LLM-A$_i$

Again, given the known variability in LLM responses, we additionally define four ensemble baselines that vote over the outputs of the four single-call outputs: at least one flags inconsistency (Voting$_{\geq1}$), at least two (Voting$_{\geq2}$), at least three (Voting$_{\geq3}$), and unanimity (Voting$_{=4}$).
These eight variations collectively capture the variability and try to mitigate the limitations of directly using LLMs for this task, to ensure a fair comparison.

We are thus left with four main approaches: Baselines with a generic multipurpose LLM, ensemble
votings over these LLMs, a code-LLM fine-tuned for inconsistency detection (\textit{C4RLLaMA}) and a
model trained from ground up for this task (\textsc{DocChecker})

\subsection{Metrics}\label{sec:metrics}
All Subjects: \tool, the different LLM-baselines, \textsc{DocChecker} as well as \textit{C4RLLaMA} predict for a given method whether it contains an inconsistency. An inconsistent method is predicted as a positive, while any method where no clear inconsistency can be determined, is predicted as a negative.
If a method truly contains an inconsistency and the tool correctly predicts it as such, it is a true positive \textbf{(TP)}. If the tool flags an actually consistent method as inconsistent, it is a false positive \textbf{(FP)}. Conversely, if a method is consistent and correctly identified as such, it is a true negative \textbf{(TN)}, and if an inconsistency is missed, it is a false negative \textbf{(FN)}.
We evaluate the different approaches using the following standard metrics for classification problems.

\begin{equation*}
    rec = \frac{TP}{TP + FN} \quad\quad  prec = \frac{TP}{TP + FP} \quad\quad spec = \frac{TN}{TN + FP} \quad \quad F_1=\frac{2\cdot prec \cdot rec}{prec + rec}
\end{equation*}

\textit{Recall} (\textit{rec.}) measures the proportion of actual inconsistencies in the project that the tool correctly identifies. High recall would be essential if the goal was to detect as many existing inconsistencies as possible. 
\textit{Precision} (\textit{prec.}) measures the proportion of samples flagged by the tool as inconsistent that are actually true inconsistencies. High precision is crucial to ensure developers are not overwhelmed with false positives.
\textit{Specificity} (\textit{spec.}) measures the proportion of consistent code-documentation pairs that are correctly identified as consistent. While it is not a standard metric in most classification evaluations, it is particularly important in our setting to ensure that \tool does not incorrectly flag too many correct methods as inconsistent.
$F_1$-score is the harmonic mean of precision and recall. It provides a single metric that balances both false positives and false negatives, making it useful when both high precision and high recall are desired.

One feature of our dataset is that it contains a core set of paired samples. For these method-documentation pairs that are inconsistent, we provide a corrected, consistent version. This structure enables us to calculate a pair-wise metric: \textit{Pair-wise Fix Precision} (\textit{PFP}). We define PFP as the fraction of true positive predictions (i.e., inconsistent versions correctly classified as inconsistent) for which the corresponding fixed version is also classified correctly as consistent.
A high \textit{PFP} indicates that the tool not only detects an inconsistency but also recognizes its resolution. In other words, if a function is correctly flagged as inconsistent, its patched counterpart is likewise classified as consistent. 

\section{Results and Evaluation}
\subsection{RQ1: Performance}
\begin{table}[tb]
   \centering
   \caption{Results of the benchmark for a balanced 50\%/50\% split (71 inconsistent/71 consistent) and an imbalanced 10\%/90\% split (71 inconsistent/639 consistent). For the imbalanced setting, medians over 1,000 randomly sampled subsets are reported. All values are rounded to two decimals.}\label{tab:results}
\begin{tabular}{@{}llrrrrrrrrrrr@{}}
\toprule
                                                                                 &                      & \multicolumn{5}{c}{50\%/50\% Split}             & \multicolumn{1}{l}{} & \multicolumn{5}{c}{10\%/90\% Split}             \\ \cmidrule(lr){3-7} \cmidrule(l){9-13} 
                                                                                 &                      & $prec.$ & $spec.$ & $rec.$ & \textit{PFP} & $F_1$ &                      & $prec.$ & $spec.$ & $rec.$ & \textit{PFP} & $F_1$ \\ \midrule
\multirow{4}{*}{\begin{tabular}[c]{@{}l@{}}general \\ LLMs\end{tabular}} & LLM-S$_c$            & .53     & .42     & .67    & .26          & .59   &                      & .15     & .58     & .67    & .26          & .24   \\
                                                                                 & LLM-S$_i$            & .56     & .58     & .54    & .37          & .55   &                      & .10     & .47     & .54    & .37          & .17   \\
                                                                                 & LLM-A$_c$            & .55     & .64     & .46    & .53          & .50   &                      & .20     & .80     & .46    & .53          & \textbf{.28}   \\
                                                                                 & LLM-A$_i$            & .59     & .81     & .29    & .65          & .39   &                      & .11     & .75     & .29    & .65          & .16   \\
\multirow{4}{*}{LLM voting}                                                      & Voting$_{\geq1}$     & .54     & .32     & .81    & .26          & \textbf{.65}   &                      & .12     & .34     & .81    & .26          & .21   \\
                                                                                 & Voting$_{\geq2}$     & .54     & .44     & .67    & .32          & .60   &                      & .14     & .53     & .67    & .32          & .23   \\
                                                                                 & Voting$_{\geq3}$     & .61     & .76     & .37    & .69          & .46   &                      & .17     & .80     & .37    & .69          & .23   \\
                                                                                 & Voting$_{=4}$        & .54     & .92     & .10    & .57          & .17   &                      & .11     & .91     & .10    & .57          & .11   \\
\multirow{2}{*}{\textsc{DocChecker}}                            & block                & .48     & .19     & .77    & .02          & .59   &                      & .08     & .06     & .77    & .02          & .15   \\
                                                                                 & single lines         & .49     & .03     & \textbf{.96}    & 0            & \textbf{.65}   &                      & .10     & .01     & \textbf{.96}    & 0            & .17   \\
\multirow{2}{*}{\textit{C4RLLaMA}}                                               & block                & .45     & .63     & .31    & .41          & .37   &                      & .12     & .75     & .31    & .41          & .18   \\
                                                                                 & single lines         & .53     & .64     & .41    & .48          & .46   &                      & .13     & .69     & .41    & .48          & .19   \\ \midrule
\multicolumn{2}{c}{\textbf{\tool}}    & \textbf{.88}     & \textbf{.97}     & .21    & \textbf{.87}          & .35   &                      & \textbf{.39}     & \textbf{.96}     & .21    &\textbf{ .87}          & \textbf{.28 }  \\ \bottomrule
\end{tabular}
\end{table}

Table~\ref{tab:results} reports precision ($prec.$), recall ($rec.$), specificity ($spec.$), pair-wise fix precision (PFP), and $F_1$-score for all configurations of the three evaluated approaches. In addition, results are presented for two different dataset splits. The first split is balanced (50\%/50\%), consisting of 71 inconsistent samples and their 71 paired consistent counterparts. The second split reflects a more realistic class distribution (10\%/90\%) of 71 inconsistent and 639 consistent samples.
For the 10\%/90\% setting, the consistent samples include the 71 paired counterparts of the inconsistent instances, as well as 568 additional samples drawn from a pool of 743 available consistent instances. We classified every sample with every tool adn version, and for the split, we generate 1,000 random subsets of consistent samples and report the median performance across runs.
It should be noted that recall and PFP remain constant across different dataset splits. Both metrics depend solely on the number of true positives. Since the set of 71 inconsistent samples is fixed and no additional true positives can be identified by increasing the number of consistent samples, these values do not change across splits.

The first notable observation is the poor performance of \textsc{DocChecker}. 
Both variants exhibit worse than random discriminative power with a precision below 50\%. \textsc{DocChecker} does achieve the highest recall of all tools for both splits. However, a value of 0.96 is reached simply by flagging almost everything as inconsistent, which is not helpful for developers. In the original paper for \textit{C4RLLaMA} a strong improvement over \textsc{DocChecker} is reported~\cite{carlama}.  Here, we can also see that it is generally performing better in many metrics, especially for the 10\%/90\% split.

Comparing the two C4RLLaMA variants, the line-by-line documentation processing (single lines) improves all metrics over the block-based version. However, a precision of .53 is still well below the results achieved by \tool or even basic, non-fine-tuned LLMs. Also notable is the consistently low \textit{Pair-wise Fix Precision} \textit{(PFP}) for both \textsc{DocChecker} and \textit{C4RLLaMA}. This suggests that these specialized models may be overspecialized on their training data. The low \textit{PFP} indicates that they tend to classify both the consistent and inconsistent version of the same method as inconsistent, revealing that they may not correctly detect inconsistencies but instead latched onto some unrelated code patterns. In contrast, \tool’s test-case-based approach should handle this problem more effectively.

Looking at the baseline LLM runs, one interesting behavior is that prompt direction affects the single‑shot LLMs more than the amount of context. Asking ``\emph{Are code and documentation consistent?}'' (LLM-S$_{c}$ and LLM-A$_{c}$) generally yields more positive answers than asking ``\emph{Is there an inconsistency?}'' (LLM-S$_{i}$ and LLM-A$_{i}$).
Since the baselines return not only a `\textit{yes}' or `\textit{no}' but also an explanation, we are able to do a manual inspection of what leads to these results. When queried for consistency, the model often focuses on general style or potentially missing things like error handling, and thus more eagerly reports an inconsistency. When queried directly for inconsistencies, it often requires a specific word or line that is wrong. This stricter threshold explains why both `inconsistency' prompts have higher precision but lower recall than the repsective `consistency' prompts. 
Adding more context in the advanced prompts (LLM-A$_{i}$ and LLM-A$_{c}$) generally shows only a small improvement accross both prompting directions and dataset splits, however, the recal drops strongly indicating that the extra information leads to less reported cases in general. Presumably because, the model sometimes gets distracted and drifts into more general issues.
All in all, no individual LLM-baseline can be trusted completely, which is why we employed voting. 

Ensembling mitigates the volatility of individual LLMs. Voting$_{\geq3}$ as the best of the four variants surpasses \tool's \textit{PFP} score (0.92 vs. 0.86) but still has lower precision and specificity. It is also interesting to see that the different baselines report different samples as inconsistent; otherwise, the voting could not return more true positives (and false positives) than any single baseline does. This again shows the flakiness of individual LLM results, and while it is possible to somewhat mitigate those issues by ensemble voting, it cannot consistently beat a more strict test-based method such as \tool.

\begin{figure}[ht]
    \centering
    \begin{subfigure}[b]{0.49\textwidth}
        \centering
        \includegraphics[width=\textwidth]{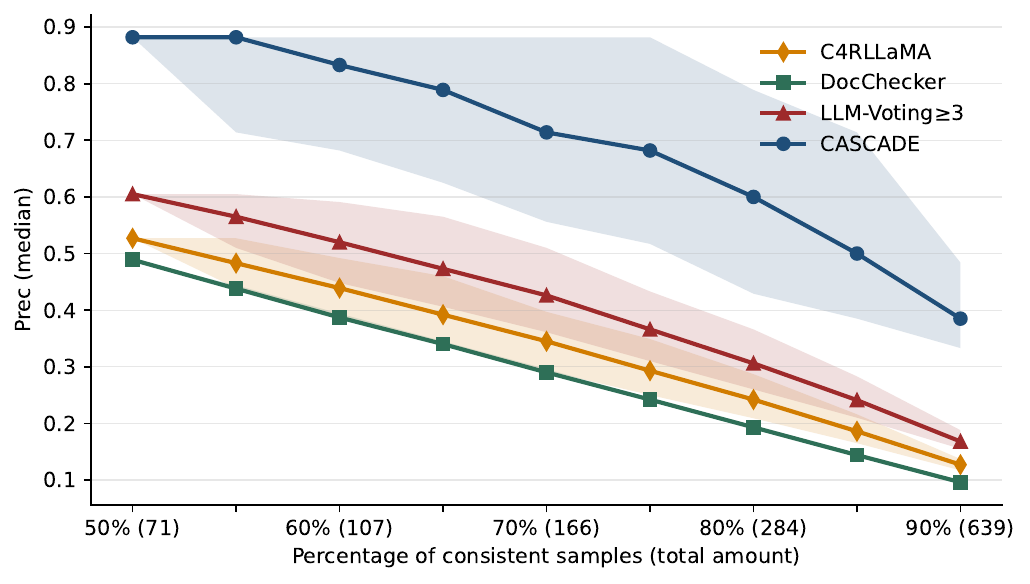}
        \caption{Precision across different dataset splits}
        \label{fig:trend:sub1}
    \end{subfigure}
    \hfill
    \begin{subfigure}[b]{0.49\textwidth}
        \centering
        \includegraphics[width=\textwidth]{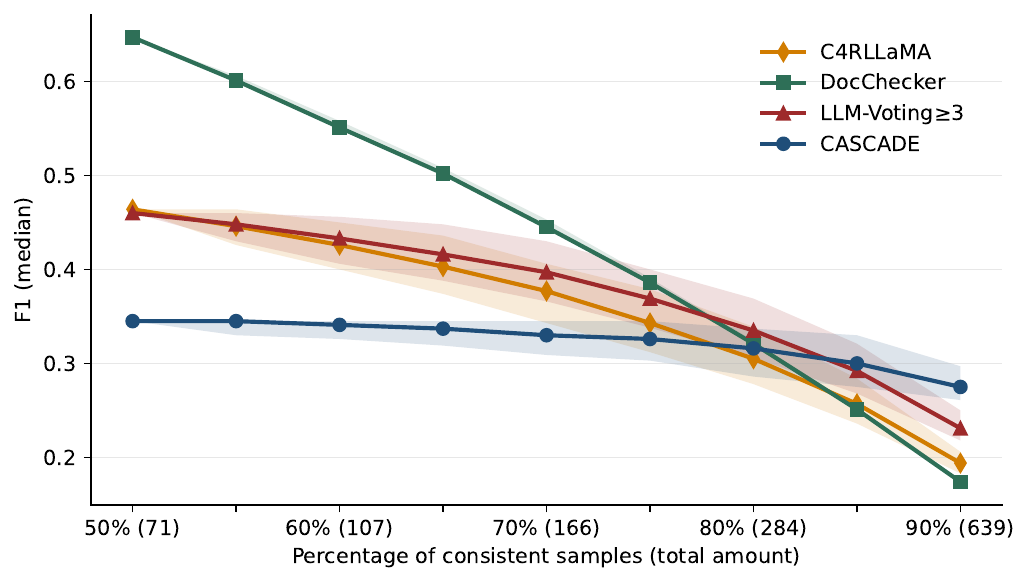}
        \caption{$F_1$-score across different dataset splits}
        \label{fig:trend:sub2}
    \end{subfigure}
    \caption{Median (line) and max and min values (shaded area) over 1,000 sampled subsets is shown. Inconsistencies stay fixed at 71. }
    \label{fig:trend}
\end{figure}

An important observation concerns stability under class imbalance. In real-world software projects, inconsistencies between documentation and implementation are typically rare, but their exact amount is unknown. Estimating it reliably would require either extremly  large-scale manual inspection or a sufficiently accurate detection approach, which we are attempting to build here. Still, evaluation under skewed class distributions is more representative of practical deployment scenarios than a purely balanced benchmark.
Figure~\ref{fig:trend} shows trends accros increasingly unbalanced dataset splits for four approaches: \tool, the best-performing LLM-based voting variant (Voting$_{\geq3}$), the best-performing \textsc{DocChecker} configuration (single-line), and the best-performing \textit{C4RLLaMA} configuration (single-line) (`best-performing' based on precision on the 50\%/50\% split). For the data displayed in Figure~\ref{fig:trend}, we fix the 71 inconsistent instances and their consistent counterparts and progressively increase the number of consistent samples to simulate different distributions.
The x-axis in both subfigures represents the proportion of inconsistent to consistent samples. The evaluation ranges from a balanced 50\%/50\% split (71 inconsistent and 71 consistent samples) to increasingly imbalanced settings, up to 10\%/90\% (71 inconsistent and 639 consistent samples).
The consistent samples are drawn from the 743 available consistent instances. For every class ratio, we generate 1,000 random subsets of consistent samples and report the median performance. The shaded area around each curve indicates the minimum and maximum values observed across these 1,000 runs.

Subfigure \ref{fig:trend:sub1} shows that precision decreases for all approaches as the dataset becomes increasingly imbalanced. This behavior is expected. While the number of true positives is bounded by the fixed set of 71 inconsistent samples, the number of false positives can increase as more consistent samples are added. Since all evaluated approaches produce false positives, precision declines monotonically across splits. \tool achieves the highest precision across all class distributions. However, it also exhibits higher variability compared to the other methods. This can be explained by the fact that \tool produces only 31 false positives across the full set of 814 consistent samples. Consequently, whether specific false positives are included or excluded in a randomly sampled subset has a relatively large impact on precision. In contrast, \textsc{DocChecker} shows almost no variability, as it consistently overreports. Its number of false positives is sufficiently large so that changes in the sampled consistent subset have negligible relative impact.

In Subfigure~\ref{fig:trend:sub2} the $F_{1}$-score is shown. Due to the precision first design and the resulting low recall, \tool performs worst on the balanced split. However, as the dataset becomes more representative of real-world conditions (i.e., increasingly skewed toward consistent samples), the $F_{1}$-scores of the other approaches drop substantially. \tool remains comparatively stable and achieves the highest $F_{1}$-score for the 10\%/90\% split. Although the absolute value remains modest (approximately 0.28), it exceeds the performance of all other methods under this more realistic class distribution.
The quantitative results in Table~\ref{tab:realworld} support this observation. Only one additional approach (LLM-A$_{c}$) achieves a comparable $F_{1}$-score; however, it exhibits lower precision, lower specificity, and a worse \textit{PFP}-rate.

\tool excels in precision and specificity, with expectedly lower recall; which aligns with our design goals. This trade-off is deliberate: the system is tuned to minimize false positives, prioritizing trust in every flagged inconsistency. 
This design rationale is further supported by the results under the imbalanced 10\%/90\% split, which more closely reflects realistic project conditions.
In this imbalanced setting, methods with higher false positive rates degrade substantially because the majority of methods are consistent. In contrast, \tool remains stable due to its conservative predictions. Although recall is limited, the low false positive rate prevents an excess of incorrect reports.

Overall, when judged on the primary objectives, \tool is generally performing best. 
LLM-voting offers a recall/\textit{PFP} trade‑off that may appeal to defect‑finding tasks under higher recall constraints but still yields flaky results, whereas the approaches based on classical machine learning either drown developers in false positives or miss subtle inconsistencies entirely.
For settings where false alarms are costly and the trustworthiness of each predicted sample is important, \tool represents the most effective choice among the evaluated techniques.

\begin{tcolorbox}[mystyle]
\textbf{Answering RQ1}: \tool flags code-documentation inconsistencies in the benchmark (Section~\ref{sec:dataset}) with a very low number of false positives, and thus outperforms the baselines in the primary objectives \emph{precision} (0.88) and \emph{specificity} (0.97) for a balanced dataset, as well as in $F_{1}$-score for a realistic dataset.
\end{tcolorbox}

\subsection{RQ2: Ablation}
\label{sec:ablation}
\begin{table}[tb]

\centering 
\caption{Results based on the different stages of \tool on the 50\%/50\% split} \label{tab:ablation} 
\begin{tabular}{@{}l@{\hskip -2pt}r@{\hskip -1pt}r@{\hskip -1pt}r@{\hskip -1pt}r@{\hskip -1pt}rc@{\hskip -2pt}r@{\hskip -1pt}r@{\hskip -1pt}r@{\hskip -1pt}r@{}}
\toprule
  & \phantom{n} \textit{prec.}  &   \phantom{n} \textit{spec.} & \phantom{n} \textit{rec.} & \phantom{n} \textit{PFP} & \phantom{n} $F_1$  &  & \phantom{n} TP & \phantom{n} FP    & \phantom{n} TN & \phantom{n} FN \\ \midrule
\textit{Phase 1}
& .60 & .63 & .59  & .54 & .59 & &41 & 27  & 45 & 29   \\
\textit{Phase 2}
& .85& .96 & .24 & .88& .38 & & 17& 3& 69& 53  \\
\textit{full }  
& .88 & .97   & .21 & .87 & .35 & & 15 & 2& 70 & 55\\\bottomrule

\end{tabular}

\end{table}

Our primary goal is to detect as many inconsistencies as possible while minimizing the number of false positives. This represents a classic trade-off in classification tasks: increasing precision often results in lower recall, and vice versa.
In \tool's design (Section \ref{sec:cascade}), we introduced multiple steps aimed at optimizing this trade-off. 
In Table~\ref{tab:ablation}, we show three different configurations of the tool with the evaluation metrics defined earlier (see Section~\ref{sec:metrics}). Namely, the prediction values from the confusion matrix (TP, FP, TN, FN), precision (\textit{prec.}), specificity (\textit{spec.}), recall (\textit{rec.}) and the \textit{Pair-wise Fix Precision} (\textit{PFP}).

Without the second phase enabled (\textit{Phase 1}), \tool reports an inconsistency as soon as \emph{any} generated test fails.  
This configuration catches 41 out of 71 real inconsistencies (recall of 0.59) but at the cost of 27 false positives, which drags precision down to 0.60 and specificity to 0.63. With this score, it would still be on par or better in terms of recall with the LLM baselines (see Table~\ref{tab:results}) and beat many of them in precision as well. However, our declared goal is to reduce the false positives burden for developers so to get rid of as many of the false positives as possible and thus we introduced \textit{Phase~2}. 

The basic version of \textit{Phase 2} predicts an inconsistency as soon as any test flips from failing to passing (\#f2p\textgreater0). This means we do not have the added check that there are no test cases that passed before and failed now. This final version (\textit{full}) eliminates most of the wrong reports, lifting precision to 0.85 and specificity to 0.96 at the cost of recall. This confirms our design intuition that the failing tests alone are not enough of an indicator because with the generated code we actually filter out wrong generated tests.

We manually evaluated the two single false positives predicted by \tool.
and they both came from cases wehre the generated code and tests made the same false asumption over something underspecified in the documentation.

False negatives, by contrast, mostly arise from tests that do not compile or from missing discriminating cases. We expect these limitations to diminish as LLM quality and test generation techniques improve in the future.

The final check, if the amount of p2f tests is 0, actually helps to sort out wrongly generated code, as intended (\tool$_{full}$). If there is at least one test case that passed before and now fails on the new code, we cannot really trust this new code on the passing test cases.
This step removes one of the two remaining false positives, slightly improving precision (0.82 to 0.88) with only a marginal decrease in recall (0.26 to 0.21). 
The full version thus eliminated all but one false positive which has been analyzed in the previous section.

An analysis of the different classification types reveals that for a given method, roughly 5-20 tests are generated (average of 8.4). The average numbers for the results classes are: p2p:5.9; f2f:2; p2f:0.3 and f2p:0.4. 
At most, three f2p (usually 1-2) cases exist which are testing exactly the one thing that is wrong in the documentation. During \tool's development, we found two main reasons for non-compiling tests: uncaught declared exceptions and the LLM hallucinating non-existent functions or fields. This will most likely improve with better base LLMs.

Overall, Phase~2 accounts for the bulk of the precision improvement by discarding spurious failures, while the additional requirement p2f=0 prunes misclassifications from wrong generated code. The resulting \tool$_{full}$ configuration leaves a single false positive and increases precision to 0.88 with a decrease in recall (0.26 to 0.21). Depending on project goals, \textit{no-Phase-2} is preferable when maximizing recall, whereas \textit{Phase-2-no-p2f=0} trades a few found cases for a substantial drop in false alarms.

\begin{tcolorbox}[mystyle]
\textbf{Answering RQ2}: \tool's different steps serve the purpose of False Positive reduction, in turn enhancing precision, at the cost of recall.
\end{tcolorbox}

\subsection{RQ3: Inconsistencies in the Wild}
During the development of \tool, we applied multiple versions of the tool to real-world Java projects. To assess the generalizability of our approach, we also developed C\# and Rust versions, adapting the tool to each language by implementing custom extractors and executors, and tailoring the prompts to generate appropriate unit tests. 
We checked the latest versions of 15 of the projects from our dataset collection for Java, 13 high starred GitHub projects for C\#, and 6 Projects from Zhang et al.~\cite{rustc4} for Rust.
We did not report very minor possible inconsistencies. 

First, some inconsistencies resulted from incomplete documentation of exceptional behavior. For example, while the documentation did not specify which exception is thrown for invalid input, the implementation raised a reasonable but different exception than assumed by \tool. Although technically inconsistent, such cases reflect underspecification rather than clear semantic contradictions and were therefore not considered substantial enough to be reported.
Second, we observed instances of undocumented behavior without observable semantic impact. For example, a method in a Stringbuilder class was documented to ignore null values but additionally explicitly ignored empty strings (""). While this behavior deviates from the specification, it does not affect the resulting observable program behavior and is therefore practically irrelevant.
To again avoid unnecessary overhead on maintainers, we restricted reporting to inconsistencies with clear semantic impact.

Through these cross-language experiments, we identified 13 previously undocumented real inconsistencies, which we reported to the respective maintainers.\footnote{
Java:
\href{https://issues.apache.org/jira/browse/LANG-173}{Commons Lang-173}, 
\href{https://issues.apache.org/jira/browse/TEXT-234}{Commons Text-234} (2 reported), 
\href{https://github.com/Atmosphere/atmosphere/issues/2514}{Atmosphere\#2514}, 
\href{https://github.com/Atmosphere/atmosphere/issues/2515 }{Atmosphere\#2515}, 
\href{https://github.com/jfree/jfreechart/issues/417}{JFreeChart\#417} (2 reported); 
C\#: 
\href{https://github.com/jellyfin/jellyfin/pull/13360}{jellyfin\#13360}, 
\href{https://github.com/bchavez/Bogus/issues/581}{Bogus\#581}, 
\href{https://github.com/icsharpcode/SharpZipLib/issues/887}{SharpZipLib\#887}; 
Rust: 
\href{https://github.com/avhz/RustQuant/pull/296}{RustQuant\#296} (2 reported),
\href{https://github.com/bitshifter/glam-rs/issues/622}{glam-rs\#622}.}  

Table~\ref{tab:realworld} summarizes the results for the real-world evaluation. The column \textit{checked} reports the number of functions analyzed. We considered only public, non-constructor, non-abstract methods with available documentation.
\textit{FP} denotes the number of produced false positives. These primarily resulted from cases in which both the generated tests and the generated code assumptions were incorrect. Such cases were identified and filtered out through manual inspection.
\textit{TP total} represents the total number of confirmed true positives. We further distinguish between \textit{minor} cases and those severe enough to warant a report. \textit{Reported} cases include the remaining confirmed inconsistencies that were communicated to project maintainers via GitHub pull-requests, GitHub issues or external bug trackers (whatever the specific project maintainers prefered) and fixed (\textit{Fix.}) denotes the number of issues that have since been resolved, either through our submitted pull requests or through independent fixes by the original developers.
In one instance, maintainers argued that the reported behavior did not constitute a genuine inconsistency. This case involved a hash function with behavior slightly deviating from the documentation but still satisfying the general Java hash contract. We therefore reclassified it as minor. The remaining unacknowledged cases were primarily in repositories with limited maintenance activity.

The observed drop in precision compared to the results of RQ1 and RQ2 can likely be attributed to two main factors. First, the real-world evaluation used an older and less capable model (\texttt{gpt-4o-mini} instead of \texttt{gpt-4.1-mini}), primarily for cost reasons, as the analyzed projects were substantially larger than the curated dataset.
Second, many real-world projects exhibit low documentation quality. Since our approach targets inconsistencies between documentation and implementation, poorly specified or low-quality documentation inherently limits the achievable precision, as it becomes difficult to determine whether deviations constitute genuine inconsistencies.

We now discuss two illustrative examples for the additinal languages, as they reveal noteworthy insights about \tool. Our running example in Figure~\ref{fig:example} is already an illustrative example of one of the Java inconsistencies we found and reported.

\begin{table}[tb]
\centering 
\caption{An overview of the real world issues found by \tool}\label{tab:realworld}
\begin{tabular}{@{}llrccccc@{}}
\toprule
\multirow{2}{*}{Language} & \multirow{2}{*}{author/project} & \multirow{2}{*}{checked} & \multicolumn{1}{c}{\multirow{2}{*}{FP}} & \multicolumn{4}{c}{TP}                                                                                                 \\ \cmidrule(l){5-8} 
                          &                                 &                          & \multicolumn{1}{c}{}                    & \multicolumn{1}{c}{Total} & \multicolumn{1}{c}{Minor} & \multicolumn{1}{c}{Reported}  & \multicolumn{1}{l}{Fix.} \\ \midrule
\multirow{4}{*}{Java}     & apache/commons-lang             & 986                      & 2                                       & 2                         & \textit{1}                & 1                                & 1                        \\
                          & apache/commons-text             & 575                      & 3                                       & 4                         & \textit{2}                & 2                                & 2                        \\
                          & Atmosphere/atmosphere           & 543                      & 3                                       & 2                         & \textit{0}                & 2                                & 2                        \\
                          & jfree/jfreechart                & 4823                     & 9                                       & 10                        & \textit{8}                & 2                                & -                        \\
\multirow{3}{*}{C\#}      & jellyfin/jellyfin               & 93                       & 7                                       & 2                         & \textit{1}                & 1                                & 1                        \\
                          & bchavez/Bogus                   & 211                      & 1                                       & 1                         & \textit{0}                & 1                                & 1                        \\
                          & icsharpcode/SharpZipLib         & 91                       & 13                                      & 4                         & \textit{3}                & 1                                & -                        \\
\multirow{2}{*}{Rust}     & avhz/RustQuant                  & 186                      & 2                                       & 2                         & \textit{0}                & 2                                & 2                        \\
                          & bitshifter/glam-rs              & 248                      & 4                                       & 3                         & \textit{2}                & 1                                & 1                        \\ \bottomrule
\end{tabular}
\end{table}

\subsubsection*{\textbf{C\#}}
For C\# we had no prior work to start with, so we looked at high starred projects on GitHub\footnote{jellyfin/jellyfin, 
bchavez/Bogus, 
icsharpcode/SharpZipLib, 
Humanizr/Humanizer, 
dotnet/aspnetcore, 
FluentValidation/FluentValidation, 
LuckyPennySoftware/AutoMapper, 
JoshClose/CsvHelper, 
BrighterCommand/Brighter, 
adamralph/bullseye, 
Tyrrrz/CliWrap, 
sipsorcery-org/sipsorcery,
thecodrr/BreadPlayer}.
\\\textbf{Bogus} is an open-source data generator for C\# that is widely used and included in several Microsoft projects. It provides a method called \texttt{ULong} that generates a random unsigned long value between a given \texttt{min} and \texttt{max} shown in Figure~\ref{fig:bogus}.
\tool discovered that the method consistently crashes with an arithmetic overflow when min is close to the max value of long, due to a floating point rounding error.
This behavior violates the expected functionality described in the documentation, which promises reliable generation of values across the full range of an unsigned long.
To reveal the problem, \tool generated tests with high \texttt{min} values, which resulted in predictable runtime exceptions. The conversion of the randomized value to a long was happening too late in the equation. With its proposed code, the tool assisted us in identifying the root cause of the problem, highlighting its potential as a valuable developer assistant.

\lstdefinelanguage{CSharp}{
    morekeywords={abstract, as, base, bool, break, byte, case, catch, char, checked, class, const, continue,
        decimal, default, delegate, do, double, else, enum, event, explicit, extern, false, finally, fixed, float,
        for, foreach, goto, if, implicit, in, int, interface, internal, is, lock, long, namespace, new, null,
        object, operator, out, override, params, private, protected, public, readonly, ref, return, sbyte,
        sealed, short, sizeof, stackalloc, static, string, struct, switch, this, throw, true, try, typeof,
        uint, ulong, unchecked, unsafe, ushort, using, virtual, void, volatile, while, async, await, var, dynamic,
        yield, KernelFunction, Description},
    sensitive=true,
    morecomment=[l]{///},
    morecomment=[s]{/*}{*/},
    morestring=[b]{"},
    morestring=[b]{@"},
    morestring=[b]{@"},
    morestring=[b]{'}
}

\begin{figure}
    \centering
    \begin{lstlisting}[language=CSharp, escapechar=§]
/// Generate a random ulong between MinValue and MaxValue.
/// Parameter - min: Min value, inclusive. Default ulong.MinValue.
/// Parameter - max: Max value, inclusive. Default ulong.MaxValue.
public ulong ULong(ulong min = ulong.MinValue,
  ulong max = ulong.MaxValue)
{ return Convert.ToUInt64(Double() * (max - min) + min§\colorbox{highlightred}{)}§; }
\end{lstlisting}
    \caption{\textbf{C\#} example: Line 6 contains the expression that causes the mistake. The conversion should only be applied to the result of the multiplication. It should not include the addition of \texttt{min}.
}
\Description{a code example}
    \label{fig:bogus}
\end{figure}

\subsubsection*{\textbf{Rust}}
We selected the Rust repositories from the same set that Zhang et al.~\cite{rustc4} used to evaluate their Rust-focused inconsistency detector. 
\\\textbf{RustQuant} is an open-source library for quantitative finance and financial modeling. \tool discovered an inconsistency in the \texttt{set\_value} function illustrated in Figure \ref{fig:rustQuant}, which updates an internal quote value and returns a double precision floating point number. While the documentation does suggest that a value is being set, it omits that the function returns the difference between the old and new values instead of the new value itself, which is usually expected for setters. To detect the inconsistency, \tool generated multiple tests that were expecting the return value to match the input value.
This is a notable case, because our approach revealed an omitted aspect of correct behavior, leading to a meaningful documentation improvement rather than identifying a blatantly incorrect claim.

\definecolor{dkgreen}{rgb}{0,0.6,0}
\definecolor{gray}{rgb}{0.5,0.5,0.5}
\definecolor{mauve}{rgb}{0.58,0,0.82}
\definecolor{GrayCodeBlock}{RGB}{241,241,241}
\definecolor{BlackText}{RGB}{110,107,94}
\definecolor{RedTypename}{RGB}{182,86,17}
\definecolor{GreenString}{RGB}{96,172,57}
\definecolor{PurpleKeyword}{RGB}{184,84,212}
\definecolor{GrayComment}{RGB}{170,170,170}
\definecolor{GoldDocumentation}{RGB}{180,165,45}
\definecolor{inconsistentRed}{RGB}{245, 0, 0}
\definecolor{DarkBlue}{RGB}{0, 0, 139} 
\lstdefinelanguage{rust}
{
    keywords={
        true,false,
        unsafe,async,await,move,
        use,pub,crate,super,self,mod, struct,enum,fn,const,static,let,mut,ref,type,impl,dyn,trait,where,as, break,continue,if,else,while,for,loop,match,return,yield,in,bool,u8,u16,u32,u64,u128,i8,i16,i32,i64,i128,char,str, Self,Option,Some,None,Result,Ok,Err,String,Box,Vec,Rc,Arc,Cell,RefCell,HashMap,BTreeMap, macro_rules
    },
    comment=[l][\color{GrayComment}\slshape]{//},
    morecomment=[s][\color{GrayComment}\slshape]{/*}{*/},
    morecomment=[l]{///},
    morecomment=[s][\color{GoldDocumentation}\slshape]{/*!}{*/},
    morecomment=[l][\color{GoldDocumentation}\slshape]{//!},
    morecomment=[s][\color{RedTypename}]{\#![}{]},
    morecomment=[s][\color{RedTypename}]{\#[}{]},
    stringstyle=\color{GreenString},
    string=[b]",
    moredelim=[is][\color{inconsistentRed}\bfseries]{@}{@}, 
    moredelim=[is][\color{DarkBlue}\bfseries]{ß}{ß} 
}

\begin{figure}[htbp]
\begin{lstlisting}[language=rust, escapechar=§]
/// Set the quote value.
pub fn set_value(&mut self, value: Option<f64>) -> f64 {
    let diff = match (&self.value, &value) {
        (Some(old_value), Some(new_value)) => new_value - old_value,
        _ => 0.0,
    };
    if diff != 0.0 { self.value = value; }
    §\colorbox{highlightred}{diff}§
}
\end{lstlisting}
\caption{\textbf{Rust} example: The return statement in line 8 causes the unexpected behavior. The documentation does not state that the difference to the old value is returned.}
\Description{a code example}
\label{fig:rustQuant}
\end{figure}

\begin{tcolorbox}[mystyle]
\textbf{Answering RQ3:} \tool is able to find 13 real-world inconsistencies in 10 different projects and across 3 programming languages (Java, C\#, Rust).
\end{tcolorbox}

\subsection{Threats to Validity}

\emph{Internal Validity.}
One threat to internal validity comes from the hand-curated benchmark: the functions, commits, and documentation snippets were selected by the authors and could embed unconscious bias. To counter this, we followed an established taxonomy of inconsistency types~\cite{wen2019large} and thoroughly disclosed our selection process and protocol in Section~\ref{sec:dataset}).

For RQ1, our re-implementation and training of \textsc{DocChecker} and \textit{C4RLLaMA} might under-perform if our reproduction were flawed. We mitigated this risk by following their exact training pipeline and evaluating on the dataset reported in their papers, with acceptable performance. Further, we adjusted the data so that \textsc{DocChecker} can actually extract and use the correct documentation for java code in two different ways; all to give it a better chance at succeeding. 

\emph{External Validity.}
A threat to external validity stems from the choice of target projects used in both the dataset and the real-world evaluations. These projects may not be representative of broader software development contexts. To mitigate this risk, we selected mature libraries that have undergone extensive development and if possible were used in the evaluation by other related approaches before us~\cite{jdoctor, rustc4}. This choice was made for both the dataset and the real-world experiments to enhance generalizability. 

A related concern is that, by construction, the benchmark contains only cases where the documentation is wrong; in the wild, the fault could also lie in the code. Our real world experiments uncovered several of these instances where the \emph{code} contained the mistake. Showing that \tool's test-generation strategy can detect both faulty documentation and buggy code.

\emph{Construct Validity.}
\tool is purposely optimized for high \emph{precision} and \emph{specificity} in order to minimize the burden for developers from too many false positives. A direct consequence of that is a lower recall ceiling. Although this trade-off causes inconsistencies to be missed, it accurately reflects the practical threshold at which developers are willing to act on tool output. Section~\ref{sec:ablation} (RQ2) shows that the precision-recall trade-off can be shifted, for example by disabling Phase~2, giving users control over this balance.

\section{Conclusion}\label{sec:conc}
In this work, we presented \tool, a novel tool for detecting semantic inconsistencies between source code and its natural-language documentation. By leveraging Large Language Models to generate both unit tests and alternative implementations from documentation, \tool employs a dual check validation strategy to substantially reduce false positives; a persistent limitation in prior tools. 
We compared \tool with LLM baselines (\texttt{gpt-4.1-mini}) and state-of-the-art tools~\cite{docchecker, carlama} on a dataset with real-world inconsistencies. In this evaluation, \tool achieves the highest \textit{precision} (0.88) and \textit{specificity} (0.97).

Our evaluation across multiple languages (Java, C\#, Rust) further underscores \tool's adaptability and applicability to modern, heterogeneous codebases. Notably, \tool surfaced 13 previously undocumented inconsistencies in widely used open-source projects, of which 10 were subsequently acknowledged by maintainers and fixed. This real-world impact highlights \tool's potential, not only as a research tool, but also as a practical assistant in software maintenance.

\subsection*{Data Availability}
\label{sec:data-availability}
The data and code of \tool is availabe here: \url{https://doi.org/10.5281/zenodo.19483108}

\clearpage
\bibliographystyle{ACM-Reference-Format}
\bibliography{references}

\end{document}